\documentclass[journal,draftclsnofoot,onecolumn,12pt]{IEEEtran}

\usepackage{amsthm,amssymb,amsmath,graphicx,multirow,color,amsfonts}%
\usepackage[update,prepend]{epstopdf}

\usepackage[latin1]{inputenc}
\usepackage{tikz}
\usepackage{bbm} % for \mathbbm{1}
\usepackage{pdfpages}
\usepackage{multirow}
\usepackage{subfig}
\usepackage{comment}
 \usepackage{graphicx}
\usepackage{multicol}

\usepackage[justification=centering]{caption}
\usepackage{textcomp}
\usepackage{psfrag}
\usepackage{arydshln}
\usepackage{url}
\usepackage{soul}
\usepackage{graphicx,color}
\usepackage[nolist]{acronym}
\usepackage{algorithm,algorithmic} %algorithme
\usepackage{mathtools,lipsum}
\usepackage{cuted}
\usepackage{balance} %balance the last page

\usepackage{amsmath}
\usepackage{cite}
\usepackage{mathrsfs}
\usepackage{color}
\usepackage{txfonts}

\usepackage{float}%fix the place of figure
\usepackage[capitalise]{cleveref}
\Crefname{equation}{Eq.\!}{Eqs.\!}
\Crefname{figure}{Fig.\!}{Figs.\!}
\Crefname{tabular}{Tab.\!}{Tabs.\!}
\Crefname{section}{Section\!}{Sections.\!}

\def\nb0{{\mathbf{0}}}
\def\nb1{{\mathbf{1}}}

\newtheorem{lemma}{Lemma}

\newtheorem{definition}{Definition}

\newtheorem{theorem}{Theorem}
\newtheorem{corollary}{Corollary}

\newenvironment{sequation}{
\begin{equation}\small}{\end{equation}
}

\begin{document}
%\pagenumbering{gobble}
\graphicspath{{./Figures/}}
	\begin{acronym}

\acro{5G-NR}{5G New Radio}
\acro{3GPP}{3rd Generation Partnership Project}
\acro{ABS}{aerial base station}
\acro{AC}{address coding}
\acro{ACF}{autocorrelation function}
\acro{ACR}{autocorrelation receiver}
\acro{ADC}{analog-to-digital converter}
\acrodef{aic}[AIC]{Analog-to-Information Converter}     
\acro{AIC}[AIC]{Akaike information criterion}
\acro{aric}[ARIC]{asymmetric restricted isometry constant}
\acro{arip}[ARIP]{asymmetric restricted isometry property}

\acro{ARQ}{Automatic Repeat Request}
\acro{AUB}{asymptotic union bound}
\acrodef{awgn}[AWGN]{Additive White Gaussian Noise}     
\acro{AWGN}{additive white Gaussian noise}

\acro{APSK}[PSK]{asymmetric PSK} 

\acro{waric}[AWRICs]{asymmetric weak restricted isometry constants}
\acro{warip}[AWRIP]{asymmetric weak restricted isometry property}
\acro{BCH}{Bose, Chaudhuri, and Hocquenghem}        
\acro{BCHC}[BCHSC]{BCH based source coding}
\acro{BEP}{bit error probability}
\acro{BFC}{block fading channel}
\acro{BG}[BG]{Bernoulli-Gaussian}
\acro{BGG}{Bernoulli-Generalized Gaussian}
\acro{BPAM}{binary pulse amplitude modulation}
\acro{BPDN}{Basis Pursuit Denoising}
\acro{BPPM}{binary pulse position modulation}
\acro{BPSK}{Binary Phase Shift Keying}
\acro{BPZF}{bandpass zonal filter}
\acro{BSC}{binary symmetric channels}              
\acro{BU}[BU]{Bernoulli-uniform}
\acro{BER}{bit error rate}
\acro{BS}{base station}
\acro{BW}{BandWidth}
\acro{BLLL}{ binary log-linear learning }

\acro{CP}{Cyclic Prefix}
\acrodef{cdf}[CDF]{cumulative distribution function}   
\acro{CDF}{Cumulative Distribution Function}
\acrodef{c.d.f.}[CDF]{cumulative distribution function}
\acro{CCDF}{complementary cumulative distribution function}
\acrodef{ccdf}[CCDF]{complementary CDF}               
\acrodef{c.c.d.f.}[CCDF]{complementary cumulative distribution function}
\acro{CD}{cooperative diversity}

\acro{CDMA}{Code Division Multiple Access}
\acro{ch.f.}{characteristic function}
\acro{CIR}{channel impulse response}
\acro{cosamp}[CoSaMP]{compressive sampling matching pursuit}
\acro{CR}{cognitive radio}
\acro{cs}[CS]{compressed sensing}                   
\acrodef{cscapital}[CS]{Compressed sensing} %will not include it in the list
\acrodef{CS}[CS]{compressed sensing}
\acro{CSI}{channel state information}
\acro{CCSDS}{consultative committee for space data systems}
\acro{CC}{convolutional coding}
\acro{Covid19}[COVID-19]{Coronavirus disease}

\acro{DAA}{detect and avoid}
\acro{DAB}{digital audio broadcasting}
\acro{DCT}{discrete cosine transform}
\acro{dft}[DFT]{discrete Fourier transform}
\acro{DR}{distortion-rate}
\acro{DS}{direct sequence}
\acro{DS-SS}{direct-sequence spread-spectrum}
\acro{DTR}{differential transmitted-reference}
\acro{DVB-H}{digital video broadcasting\,--\,handheld}
\acro{DVB-T}{digital video broadcasting\,--\,terrestrial}
\acro{DL}{DownLink}
\acro{DSSS}{Direct Sequence Spread Spectrum}
\acro{DFT-s-OFDM}{Discrete Fourier Transform-spread-Orthogonal Frequency Division Multiplexing}
\acro{DAS}{Distributed Antenna System}
\acro{DNA}{DeoxyriboNucleic Acid}

\acro{EC}{European Commission}
\acro{EED}[EED]{exact eigenvalues distribution}
\acro{EIRP}{Equivalent Isotropically Radiated Power}
\acro{ELP}{equivalent low-pass}
\acro{eMBB}{Enhanced Mobile Broadband}
\acro{EMF}{ElectroMagnetic Field}
\acro{EU}{European union}
\acro{EI}{Exposure Index}
\acro{eICIC}{enhanced Inter-Cell Interference Coordination}

\acro{FC}[FC]{fusion center}
\acro{FCC}{Federal Communications Commission}
\acro{FEC}{forward error correction}
\acro{FFT}{fast Fourier transform}
\acro{FH}{frequency-hopping}
\acro{FH-SS}{frequency-hopping spread-spectrum}
\acrodef{FS}{Frame synchronization}
\acro{FSsmall}[FS]{frame synchronization}  
\acro{FDMA}{Frequency Division Multiple Access}

\acro{GA}{Gaussian approximation}
\acro{GF}{Galois field }
\acro{GG}{Generalized-Gaussian}
\acro{GIC}[GIC]{generalized information criterion}
\acro{GLRT}{generalized likelihood ratio test}
\acro{GPS}{Global Positioning System}
\acro{GMSK}{Gaussian Minimum Shift Keying}
\acro{GSMA}{Global System for Mobile communications Association}
\acro{GS}{ground station}
\acro{GMG}{ Grid-connected MicroGeneration}

\acro{HAP}{high altitude platform}
\acro{HetNet}{Heterogeneous network}

\acro{IDR}{information distortion-rate}
\acro{IFFT}{inverse fast Fourier transform}
\acro{iht}[IHT]{iterative hard thresholding}
\acro{i.i.d.}{independent, identically distributed}
\acro{IoT}{Internet of Things}                      
\acro{IR}{impulse radio}
\acro{lric}[LRIC]{lower restricted isometry constant}
\acro{lrict}[LRICt]{lower restricted isometry constant threshold}
\acro{ISI}{intersymbol interference}
\acro{ITU}{International Telecommunication Union}
\acro{ICNIRP}{International Commission on Non-Ionizing Radiation Protection}
\acro{IEEE}{Institute of Electrical and Electronics Engineers}
\acro{ICES}{IEEE international committee on electromagnetic safety}
\acro{IEC}{International Electrotechnical Commission}
\acro{IARC}{International Agency on Research on Cancer}
\acro{IS-95}{Interim Standard 95}

\acro{KPI}{Key Performance Indicator}

\acro{LEO}{low earth orbit}
\acro{LF}{likelihood function}
\acro{LLF}{log-likelihood function}
\acro{LLR}{log-likelihood ratio}
\acro{LLRT}{log-likelihood ratio test}
\acro{LoS}{Line-of-Sight}
\acro{LRT}{likelihood ratio test}
\acro{wlric}[LWRIC]{lower weak restricted isometry constant}
\acro{wlrict}[LWRICt]{LWRIC threshold}
\acro{LPWAN}{Low Power Wide Area Network}
\acro{LoRaWAN}{Low power long Range Wide Area Network}
\acro{NLoS}{Non-Line-of-Sight}
\acro{LiFi}[Li-Fi]{light-fidelity}
 \acro{LED}{light emitting diode}
 \acro{LABS}{LoS transmission with each ABS}
 \acro{NLABS}{NLoS transmission with each ABS}

\acro{MB}{multiband}
\acro{MC}{macro cell}
\acro{MDS}{mixed distributed source}
\acro{MF}{matched filter}
\acro{m.g.f.}{moment generating function}
\acro{MI}{mutual information}
\acro{MIMO}{Multiple-Input Multiple-Output}
\acro{MISO}{multiple-input single-output}
\acrodef{maxs}[MJSO]{maximum joint support cardinality}                       
\acro{ML}[ML]{maximum likelihood}
\acro{MMSE}{minimum mean-square error}
\acro{MMV}{multiple measurement vectors}
\acrodef{MOS}{model order selection}
\acro{M-PSK}[${M}$-PSK]{$M$-ary phase shift keying}                       
\acro{M-APSK}[${M}$-PSK]{$M$-ary asymmetric PSK} 
\acro{MP}{ multi-period}
\acro{MINLP}{mixed integer non-linear programming}

\acro{M-QAM}[$M$-QAM]{$M$-ary quadrature amplitude modulation}
\acro{MRC}{maximal ratio combiner}                  
\acro{maxs}[MSO]{maximum sparsity order}                                      
\acro{M2M}{Machine-to-Machine}                                                
\acro{MUI}{multi-user interference}
\acro{mMTC}{massive Machine Type Communications}      
\acro{mm-Wave}{millimeter-wave}
\acro{MP}{mobile phone}
\acro{MPE}{maximum permissible exposure}
\acro{MAC}{media access control}
\acro{NB}{narrowband}
\acro{NBI}{narrowband interference}
\acro{NLA}{nonlinear sparse approximation}
\acro{NLOS}{Non-Line of Sight}
\acro{NTIA}{National Telecommunications and Information Administration}
\acro{NTP}{National Toxicology Program}
\acro{NHS}{National Health Service}

\acro{LOS}{Line of Sight}

\acro{OC}{optimum combining}                             
\acro{OC}{optimum combining}
\acro{ODE}{operational distortion-energy}
\acro{ODR}{operational distortion-rate}
\acro{OFDM}{Orthogonal Frequency-Division Multiplexing}
\acro{omp}[OMP]{orthogonal matching pursuit}
\acro{OSMP}[OSMP]{orthogonal subspace matching pursuit}
\acro{OQAM}{offset quadrature amplitude modulation}
\acro{OQPSK}{offset QPSK}
\acro{OFDMA}{Orthogonal Frequency-division Multiple Access}
\acro{OPEX}{Operating Expenditures}
\acro{OQPSK/PM}{OQPSK with phase modulation}

\acro{PAM}{pulse amplitude modulation}
\acro{PAR}{peak-to-average ratio}
\acrodef{pdf}[PDF]{probability density function}                      
\acro{PDF}{probability density function}
\acrodef{p.d.f.}[PDF]{probability distribution function}
\acro{PDP}{power dispersion profile}
\acro{PMF}{probability mass function}                             
\acrodef{p.m.f.}[PMF]{probability mass function}
\acro{PN}{pseudo-noise}
\acro{PPM}{pulse position modulation}
\acro{PRake}{Partial Rake}
\acro{PSD}{power spectral density}
\acro{PSEP}{pairwise synchronization error probability}
\acro{PSK}{phase shift keying}
\acro{PD}{power density}
\acro{8-PSK}[$8$-PSK]{$8$-phase shift keying}
\acro{PPP}{Poisson point process}
\acro{PCP}{Poisson cluster process}
 
\acro{FSK}{Frequency Shift Keying}

\acro{QAM}{Quadrature Amplitude Modulation}
\acro{QPSK}{Quadrature Phase Shift Keying}
\acro{OQPSK/PM}{OQPSK with phase modulator }

\acro{RD}[RD]{raw data}
\acro{RDL}{"random data limit"}
\acro{ric}[RIC]{restricted isometry constant}
\acro{rict}[RICt]{restricted isometry constant threshold}
\acro{rip}[RIP]{restricted isometry property}
\acro{ROC}{receiver operating characteristic}
\acro{rq}[RQ]{Raleigh quotient}
\acro{RS}[RS]{Reed-Solomon}
\acro{RSC}[RSSC]{RS based source coding}
\acro{r.v.}{random variable}                               
\acro{R.V.}{random vector}
\acro{RMS}{root mean square}
\acro{RFR}{radiofrequency radiation}
\acro{RIS}{Reconfigurable Intelligent Surface}
\acro{RNA}{RiboNucleic Acid}
\acro{RRM}{Radio Resource Management}
\acro{RUE}{reference user equipments}
\acro{RAT}{radio access technology}
\acro{RB}{resource block}

\acro{SA}[SA-Music]{subspace-augmented MUSIC with OSMP}
\acro{SC}{small cell}
\acro{SCBSES}[SCBSES]{Source Compression Based Syndrome Encoding Scheme}
\acro{SCM}{sample covariance matrix}
\acro{SEP}{symbol error probability}
\acro{SG}[SG]{sparse-land Gaussian model}
\acro{SIMO}{single-input multiple-output}
\acro{SINR}{signal-to-interference plus noise ratio}
\acro{SIR}{signal-to-interference ratio}
\acro{SISO}{Single-Input Single-Output}
\acro{SMV}{single measurement vector}
\acro{SNR}[\textrm{SNR}]{signal-to-noise ratio} 
\acro{sp}[SP]{subspace pursuit}
\acro{SS}{spread spectrum}
\acro{SW}{sync word}
\acro{SAR}{specific absorption rate}
\acro{SSB}{synchronization signal block}
\acro{SR}{shrink and realign}

\acro{tUAV}{tethered Unmanned Aerial Vehicle}
\acro{TBS}{terrestrial base station}

\acro{uUAV}{untethered Unmanned Aerial Vehicle}
\acro{PDF}{probability density functions}

\acro{PL}{path-loss}

\acro{TH}{time-hopping}
\acro{ToA}{time-of-arrival}
\acro{TR}{transmitted-reference}
\acro{TW}{Tracy-Widom}
\acro{TWDT}{TW Distribution Tail}
\acro{TCM}{trellis coded modulation}
\acro{TDD}{Time-Division Duplexing}
\acro{TDMA}{Time Division Multiple Access}
\acro{Tx}{average transmit}

\acro{UAV}{Unmanned Aerial Vehicle}
\acro{uric}[URIC]{upper restricted isometry constant}
\acro{urict}[URICt]{upper restricted isometry constant threshold}
\acro{UWB}{ultrawide band}
\acro{UWBcap}[UWB]{Ultrawide band}   
\acro{URLLC}{Ultra Reliable Low Latency Communications}
         
\acro{wuric}[UWRIC]{upper weak restricted isometry constant}
\acro{wurict}[UWRICt]{UWRIC threshold}                
\acro{UE}{User Equipment}
\acro{UL}{UpLink}

\acro{WiM}[WiM]{weigh-in-motion}
\acro{WLAN}{wireless local area network}
\acro{wm}[WM]{Wishart matrix}                               
\acroplural{wm}[WM]{Wishart matrices}
\acro{WMAN}{wireless metropolitan area network}
\acro{WPAN}{wireless personal area network}
\acro{wric}[WRIC]{weak restricted isometry constant}
\acro{wrict}[WRICt]{weak restricted isometry constant thresholds}
\acro{wrip}[WRIP]{weak restricted isometry property}
\acro{WSN}{wireless sensor network}                        
\acro{WSS}{Wide-Sense Stationary}
\acro{WHO}{World Health Organization}
\acro{Wi-Fi}{Wireless Fidelity}

\acro{sss}[SpaSoSEnc]{sparse source syndrome encoding}

\acro{VLC}{Visible Light Communication}
\acro{VPN}{Virtual Private Network} 
\acro{RF}{Radio Frequency}
\acro{FSO}{Free Space Optics}
\acro{IoST}{Internet of Space Things}

\acro{GSM}{Global System for Mobile Communications}
\acro{2G}{Second-generation cellular network}
\acro{3G}{Third-generation cellular network}
\acro{4G}{Fourth-generation cellular network}
\acro{5G}{Fifth-generation cellular network}	
\acro{gNB}{next-generation Node-B Base Station}
\acro{NR}{New Radio}
\acro{UMTS}{Universal Mobile Telecommunications Service}
\acro{LTE}{Long Term Evolution}

\acro{QoS}{Quality of Service}
\end{acronym}
	
	%% EMF definitions
\newcommand{\SAR} {\mathrm{SAR}}
\newcommand{\WBSAR} {\mathrm{SAR}_{\mathsf{WB}}}
\newcommand{\gSAR} {\mathrm{SAR}_{10\si{\gram}}}
\newcommand{\Sab} {S_{\mathsf{ab}}}
\newcommand{\Eavg} {E_{\mathsf{avg}}}
\newcommand{\ft}{f_{\textsf{th}}}
\newcommand{\alphatf}{\alpha_{24}}

\title{
Conditional Contact Angle Distribution in LEO Satellite-Relayed Transmission
}
\author{
Ruibo Wang, Anna Talgat, {\em Student Member, IEEE} Mustafa A. Kishk, {\em Member, IEEE} and Mohamed-Slim Alouini, {\em Fellow, IEEE}
\thanks{Ruibo Wang, Anna Talgat and Mohamed-Slim Alouini are with King Abdullah University of Science and Technology (KAUST), CEMSE division, Thuwal 23955-6900, Saudi Arabia. Mustafa A. Kishk is with the Department of Electronic Engineering, National University of Ireland, Maynooth, W23 F2H6, Ireland. (e-mail: ruibo.wang@kaust.edu.sa; anna.talgat@kaust.edu.sa; mustafa.kishk@mu.ie;
slim.alouini@kaust.edu.sa). 
}
\vspace{-10mm}
}
\maketitle

\begin{abstract}
This letter characterizes the contact angle distribution based on the condition that the relay low earth orbit (LEO) satellite is in the communication range of both the ground transmitter and the ground receiver. As one of the core distributions in stochastic geometry-based routing analysis, the analytical expression of the \ac{CDF} of the conditional contact angle is derived. Furthermore, the conditional contact angle is applied to analyze the inaccessibility of common satellites between the ground transmitter and receiver. Finally, with the help of the conditional contact angle, coverage probability and achievable data rate in LEO satellite-relayed transmission are studied.
\end{abstract}


\section{Introduction}
In recent years, the explosive growth of the number of LEO satellites has opened the door for many opportunities and challenges for the development of non-terrestrial networks \cite{kodheli2020satellite}, \cite{c2014system}. Because of the significant increase in the information-carrying capacity of the satellite network, more ground communications services can be transferred to space \cite{saeed2020cubesat}. In the face of low latency and long-distance communication requirements, LEO satellites play an important role \cite{wang2022ultra}. In a ground-satellite-ground transmission, the ground transmitter needs to associate with a reliable satellite as a relay. Compared with a high orbit satellite, one of the challenges a LEO satellite has to face is its relatively small coverage \cite{vatalaro1995analysis}. In addition, the interference from massive satellite constellations further decreases the communication region. Therefore, in addition to providing stronger power \cite{okati2022nonhomogeneous}, the associated relay satellite should also locate within the reliable communication range of both the ground transmitter and receiver. The distribution of the distance between a strictly selected relay satellite and the ground transmitter is challenging but also meaningful. In relay transmission and routing problems \cite{routingimportant}, this distribution is the basis of many system parameter analyses \cite{andrews2016primer}.
\par

Several related studies have appeared in recent years. They provide effective mathematical methods, accurate models, and practical methods for deriving the distance distribution \cite{al2021session,wang2022ultra,talgat2020nearest}. Traditional deterministic network models that extend from cellular networks (such as the spherical Voronoi model) are not be suitable for large-scale dynamic network modeling. In addition to limiting the user's distance under single-hop communication, the spherical Voronoi model is not accurate to study the distance distribution \cite{kak2019large}. Compared to the above modeling methods, the stochastic geometry-based method is undoubtedly more practical for irregular network modeling \cite{haenggi2012stochastic}. Among the existing stochastic geometry models, binomial point process (BPP) is relatively accurate for closed area networks with a fixed number of satellites \cite{talgat2020stochastic,ok-1}. Some studies have given different forms of distance distribution between the ground receiver and the nearest satellite \cite{okati2022nonhomogeneous,ok-1,talgat2020nearest}. Since the satellites are distributed on the sphere, using angle to express the contact distance distribution is more concise. Therefore, contact angle distribution is introduced to analyze the coverage probability and latency of satellite networks \cite{Al-1,wang2022stochastic}. Based on the existing research, the contributions of this letter are summarized as follows.
\begin{itemize}
    \item We derive an analytical expression of the contact angle distribution under the condition that the satellite is within the transmitter and receiver's communication range and its accuracy is verified.
    \item The influences of the number of satellites and the distance between the transmitter and receiver on the conditional contact angle are studied. 
    \item Based on the conditional contact angle distribution, the satellite inaccessibility in single relay routing is analyzed and extended to multiple relays routing.
    \item We explain how to obtain coverage probability and achievable data rate of uplink from transmitter to relay satellite in routing by applying conditional contact angle.
\end{itemize}

\begin{figure}[t]
	\centering
	\includegraphics[width=0.6\linewidth]{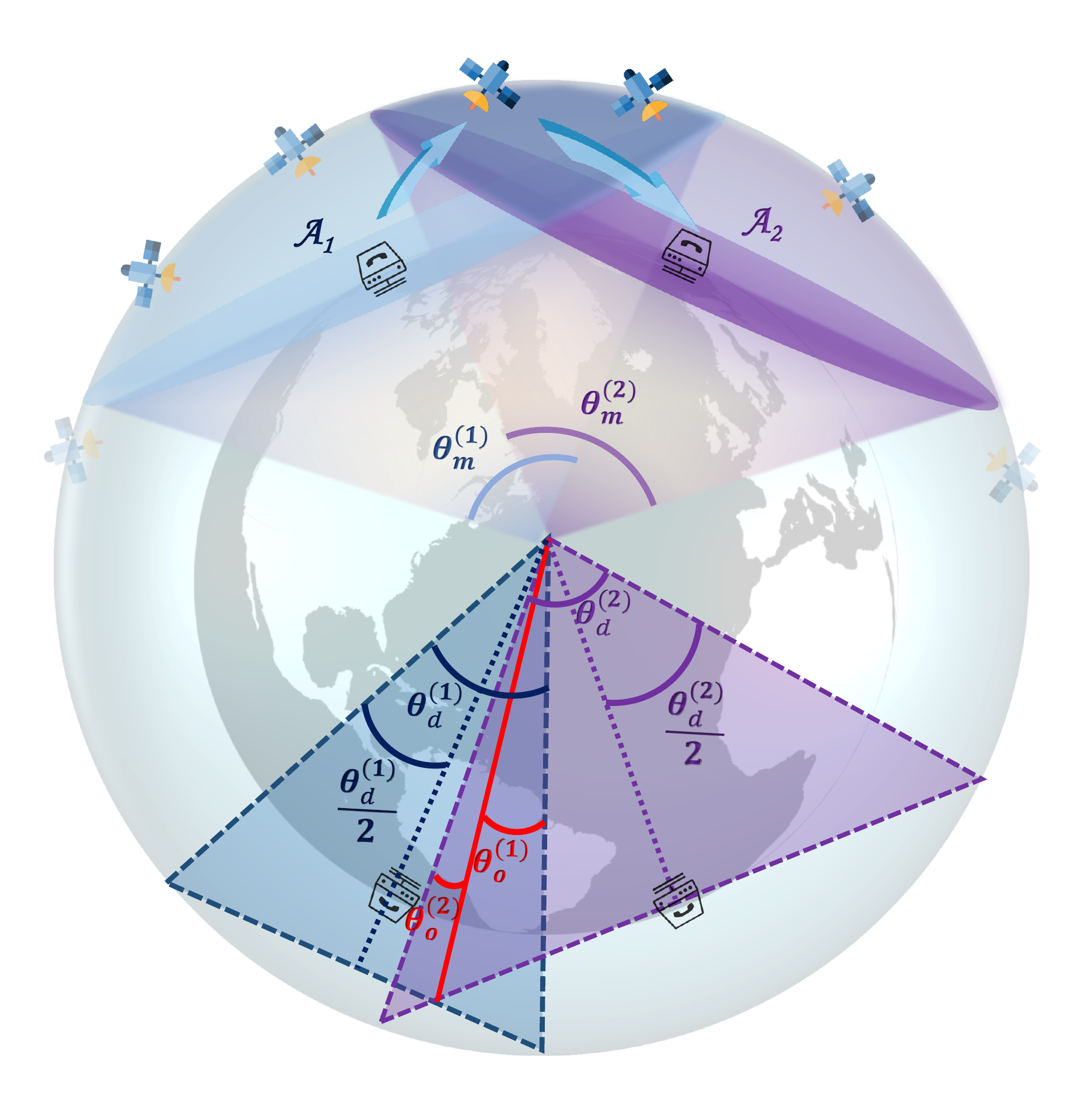}
	\caption{Satellite-relayed communication model.}
	\label{fig:Satellite-relayed communication model}
	\vspace{-0.4cm}
\end{figure}

\section{System Model}\label{system model}
In this section, we build a ground-satellite-ground relay communication model. $N_{\rm{Sat}}$ satellites are distributed on a spherical surface with radius $R_{\rm{Sat}}$ and form a homogeneous BPP \cite{ok-1}. The transmitter and receiver are located on the Earth with distance $d$. The radius of the Earth is denoted as $R_\oplus=6371\rm{km}$. We start with a simple scenario where a single satellite is selected as a relay, because the results obtained from a single satellite relay routing can be easily extended to a multiple one. A concrete example of the extension of satellite inaccessibility is provided in subsection~\ref{multiple}. 
\par
We consider a satellite that maximizes the minimum quality of service of transmitter-satellite, and satellite-receiver links is suggested to be selected. Whether the data rate, coverage probability, or latency are negatively correlated with the communication distance, the quality of service can be measured by the reciprocal of the communication distance. However, when the relay satellites form a BPP, analysis of this strategy is intractable. Therefore, we consider a slightly suboptimal but tractable selection strategy: the transmitter is assumed to choose the closest satellite as a relay among the satellites that can provide reliable communication for both the transmitter and receiver \cite{belbase2018coverage}.

\begin{definition}[Dome Angle]
Connect the two points with the center of the Earth, respectively, and the angle between the two connected lines is called the dome angle of two points.
\end{definition}

Due to Earth blockage and maximum reliable communication distance, the transmitter and receiver can only communicate with satellites within a certain area. The certain area is a spherical cap which is the shaded area ($\mathcal{A}_1$ for transmitter, and $\mathcal{A}_2$ for receiver) in the top half of Fig.~\ref{fig:Satellite-relayed communication model}. The maximum dome angles of any two points in the spherical cap are denoted as $\theta_{m}^{(1)}$ and $\theta_{m}^{(2)}$, respectively, which are also called the maximum dome angles of two spherical caps $\mathcal{A}_1$ and $\mathcal{A}_2$. To ensure that $\mathcal{A}_1$ and $\mathcal{A}_2$ have intersecting region, the following equation should be satisfied
\begin{equation}
    \theta_{m}^{(1)}+\theta_{m}^{(2)}>4\arcsin \left( \frac{d}{2R_{\oplus}} \right).
\end{equation}

\section{Conditional Contact Angle Distribution}
In this section, we derive the analytical expression of the \ac{CDF} of the conditional contact angle distribution. The definition of the conditional contact angle is given below.
\begin{definition}[Conditional Contact Angle]
Among the satellites that can provide reliable communication for the receiver, the dome angle between the transmitter and its nearest satellite is called the conditional contact angle.
\end{definition}

To derive the distribution of the conditional contact angle $\theta_c$, 
the following steps are taken: ({\romannumeral1}) fixed the region $\mathcal{A}_2$ and dome angle $\theta_d^{(2)}$, continuously increase the dome angle $\theta_d^{(1)}$ of the spherical cap corresponding to the transmitter, ({\romannumeral2}) calculate the intersecting area of the two spherical caps, and ({\romannumeral3}) calculate the probability that there are no satellites in the intersection region.
\par
To get the the intersecting area in step ({\romannumeral2}), the irregular intersecting area can be divided into two parts by the red line, shown at the bottom of Fig.~\ref{fig:Satellite-relayed communication model}. These two parts correspond to the right part of the left spherical cap and the left part of the right spherical cap. The following lemma can obtain the area of both regions. The two dome angles involved in the lemma are marked at the bottom of Fig.~\ref{fig:Satellite-relayed communication model}.

\begin{figure}[t]
	\centering
	\includegraphics[width=0.8\linewidth]{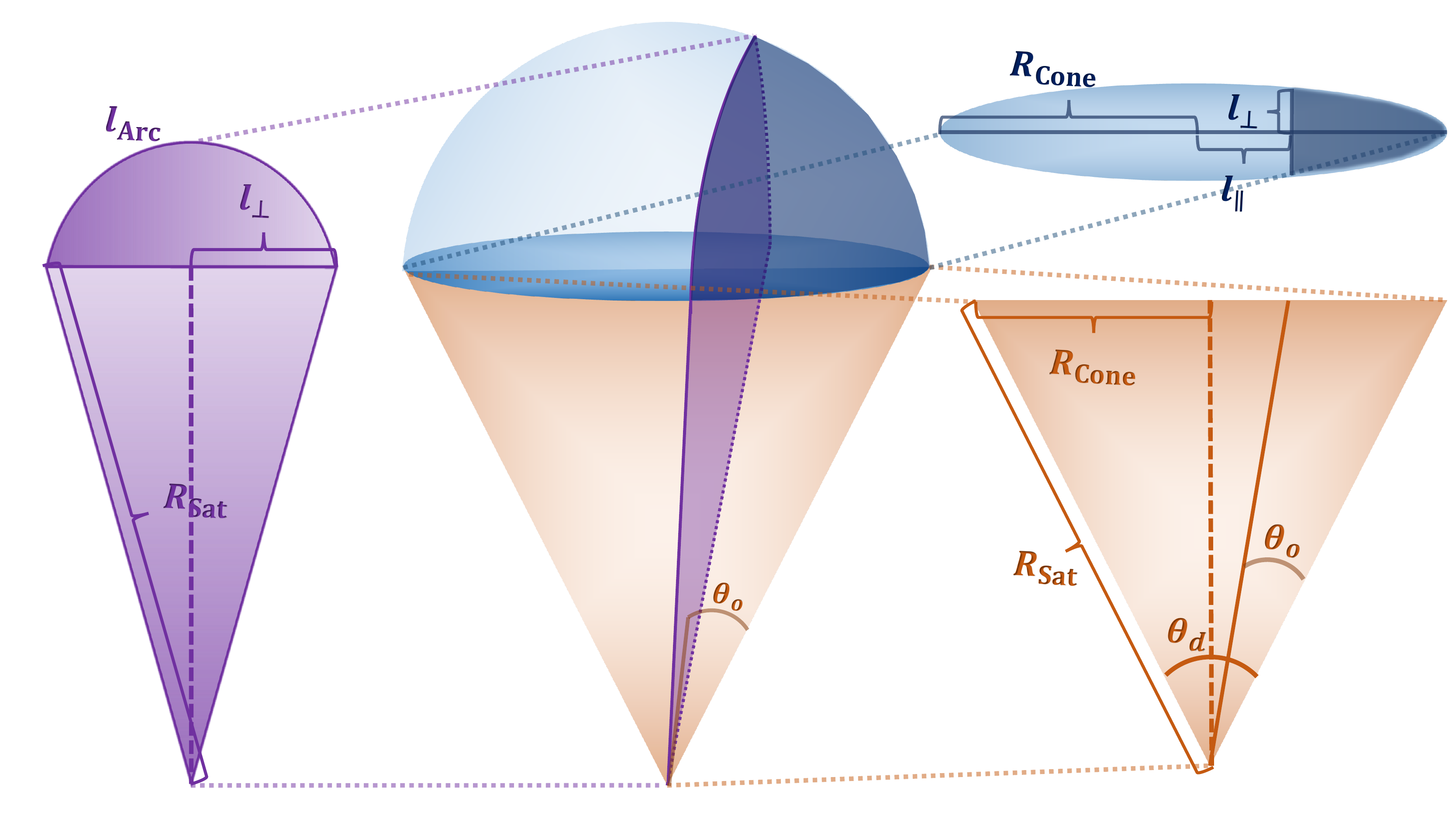}
	\caption{Decomposition diagrams of the geometry combined by a cone and spherical cap.}
	\label{fig:Decomposition diagrams of the geometry combined by a cone and cap}
	\vspace{-0.4cm}
\end{figure}

\begin{lemma}\label{area}
For a spherical cap with dome angle $\theta_d$, intercept the portion of one side of the cap, with dome angle $\theta_o$. The area of the cut part $S\left(\theta_d,\theta_o\right)$ is given by,
\begin{sequation}\label{S_xitad_xitao}
\begin{split}
    S\left(\theta_d,\theta_o\right)&=\int_{{R_{\rm{Sat}}\cos\frac{\theta_d}{2}}\tan\left(\frac{\theta_d}{2}-\theta_o\right)}^{R_{\rm{Sat}}\sin\left(\frac{\theta_d}{2}\right)}2{R_{\rm{Sat}}} \arcsin\left(\frac{\sqrt{{R_{\rm{Sat}}^2\sin\left(\frac{\theta_d}{2}\right)}^2-l^2}}{R_{\rm{Sat}}}\right)\mathrm{d}l.
\end{split}
\end{sequation}
\begin{proof}
See appendix~\ref{app:area}.
\end{proof}
\end{lemma}

Based on lemma~\ref{area}, the \ac{CDF} of the conditional contact distance is given in the following theorem.

\begin{theorem}\label{CDF of conditional contact angle}
Given that the maximum dome angle of the receiver's spherical cap is $\theta_m^{(2)}$, the approximate CDF of the conditional contact angle $F_{\theta_c}\left(\theta\right)$ is given by,
\begin{equation}\label{CDF theorem}
\begin{split}
    F_{\theta_c}\left(\theta\right) = 1 - \left(1-\frac{S\left(2\theta,\theta_o^{(1)}\left(\theta\right)\right)+S\left(\theta_{m}^{(2)},\theta_o^{(2)}\left(\theta\right)\right)}{4\pi R_{\rm{Sat}}^2}\right)^{N_{\rm{Sat}}},
\end{split}
\end{equation}
where $\theta_o^{(1)}\left(\theta\right)$ and $\theta_o^{(2)}\left(\theta\right)$ are defined as,
\begin{equation}
\begin{split}
    \theta_o^{(1)}\left(\theta\right) & = \theta - \frac{a(\theta)c-\sqrt{2a(\theta)^2-4a(\theta)b+2b^2+a(\theta)bc^2}}{a(\theta)-b}, \\
    \theta_o^{(2)}\left(\theta\right) & = \frac{1}{2}\theta_m^{(2)} - \frac{-bc+\sqrt{2a(\theta)^2-4a(\theta)b+2b^2+a(\theta)bc^2}}{a(\theta)-b},
\end{split}
\end{equation}
where
\begin{equation}
\begin{split}
    a(\theta) = \cos\theta, \ b &= \cos\frac{\theta_m^{(2)}}{2}, \ c = 2\arcsin\left(\frac{d}{2R_{\oplus}}\right),
\end{split}
\end{equation}
and the domain of the contact angle is,
\begin{equation}
\begin{split}
    \max \bigg\{0 , 2\arcsin & \left( \frac{d}{2R_{\oplus}} \right) - \frac{1}{2}\theta_{m}^{(2)} \bigg\} \leq \theta_c \\
    &\leq \min \bigg\{  \frac{1}{2} \theta_{m}^{(1)} , 2\arcsin \left( \frac{d}{2R_{\oplus}} \right)+\frac{1}{2}\theta_{m}^{(2)} \bigg\}.
\end{split}
\end{equation}

\begin{proof}
See appendix~\ref{app:CDF of conditional contact angle}
\end{proof}
\end{theorem}

\section{Potential Applications}
\subsection{Satellite Inaccessibility in Single Relay Routing}
One of the key issues in real-time and ultra long-distance routing is how far the distance between the transmitter and receiver can be or how many satellites are required to ensure that the routing will not be interrupted due to no available satellites. From theorem~\ref{CDF of conditional contact angle}, an intuitive 
 corollary about LEO relay outage probability can be obtained. The definition of LEO relay outage probability is given as follows.

\begin{definition}[LEO Relay Outage Probability]
The LEO relay outage probability is defined as the probability that there are no available satellites located in the communication range of both ground transmitter and receiver.
\end{definition}

\begin{corollary}\label{Probability Error}
Given that the Euclidean distance between the transmitter and receiver is $d$, the LEO relay outage probability $P_e^S \left(d\right) $ is given by,
\begin{equation}\label{single outage probability}
\begin{split}
   P_e^S \left(d\right) = 1 - F_{\theta_c} \bigg(\min \bigg\{  \frac{1}{2} \theta_{m}^{(1)} , 2\arcsin \left( \frac{d}{2R_{\oplus}} \right)+\frac{1}{2}\theta_{m}^{(2)} \bigg\}\bigg),
\end{split}
\end{equation}
where the CDF of the conditional contact angle $F_{\theta_c}\left(\theta\right)$ is defined in (\ref{CDF theorem}).
\end{corollary}

\subsection{Satellite Inaccessibility in Multiple Relays Routing}\label{multiple}
With the above corollary, designers can keep the outage probability below a threshold by adjusting $d$ and $N_{\rm{Sat}}$. However, only studying the satellite inaccessibility in single satellite relay routing is quite limited. The transmitter requires multiple relay satellites to send the message to the receiver in most cases. Therefore, we consider the satellite inaccessibility in a routing consisting of multiple satellite relays with a bent pipe architecture \cite{gaber20205g}.
\par
In such an architecture, $N_h - 1$ terrestrial relays are required when $N_h$ relay satellites are selected. In order to reduce latency and power consumption, we assume that every two adjacent terrestrial relays have the same dome angle. Therefore, the multiple relays outage probability $P_e^M \left(N_h,d\right) $, which is the probability that there are no satellites available in any hop, is given in the following corollary. Since $P_e^M \left(N_h,d\right) $ can be derived from $P_e^S \left(d\right) $ by simple geometric relations, the proof is omit here.
\par
\begin{corollary}
Given that the Euclidean distance between the transmitter and receiver is $d$, the multiple relays outage probability $P_e^M \left(d\right) $ is expressed as,
\begin{sequation}\label{multiple outage probability}
\begin{split}
   P_e^M \left(N_h,d\right) = 1 - \left( 1 - P_e^S \left( 2R_{\oplus} \sin \left( \frac{1}{N_h} \arcsin\left( \frac{d}{2R_{\oplus}} \right) \right) \right) \right)^{N_h},
\end{split}
\end{sequation}
where $P_e^S \left(d\right)$ is defined in (\ref{single outage probability}) and $N_h$ is the number of selected relay satellites.
\end{corollary}
\par
The $N_h$ needs to be carefully designed according to the multiple relays outage probability. Generally speaking, a long-hop strategy (with a small $N_h$) leads to a larger outage probability, while a short-hop strategy (with a large $N_h$) leads to a larger latency \cite{wang2022stochastic}.

\subsection{Uplink Coverage and Rate Analysis with Suboptimal Relay Selection}
Coverage probability and achievable maximum data rate are highlighted metrics of satellite network analysis. In the case of satellites providing coverage to ground users, the authors in \cite{talgat2020stochastic} and \cite{ok-1} provide analytical expressions of downlink coverage probability and achievable data rate, respectively. The above expressions can be modified into uplink coverage probability and achievable data rate from transmitter to satellite in the routing scenario. The modification process is straightforward, and the only task is replacing the contact distance with the conditional contact distance. Notice that the domain of contact distance also needs to be replaced. Conditional contact distance is defined as the distance from the transmitter to the closest satellite that can provide reliable communication for both the transmitter and receiver. The relationship between the conditional contact distance $d_c$ and conditional contact angle $\theta_c$ is,
\begin{equation}
    2 R_{\oplus} R_{\rm{Sat}} \cos \theta_c = R_{\oplus}^2 + R_{\rm{Sat}}^2 - d_c^2.
\end{equation}

%In addition to the multi-hop routing scenario, cross-layer communication is also a possible application scenario. The transmitter and receiver are not necessarily located on the ground, as long as they are below satellite altitude. As the dome angle of the communication region is adjustable, equipment with a height above the ground can be equivalent to the ground transmitter or receiver.

\section{Numerical Results}
\begin{figure}[t]
	\centering
	\includegraphics[width=0.6\linewidth]{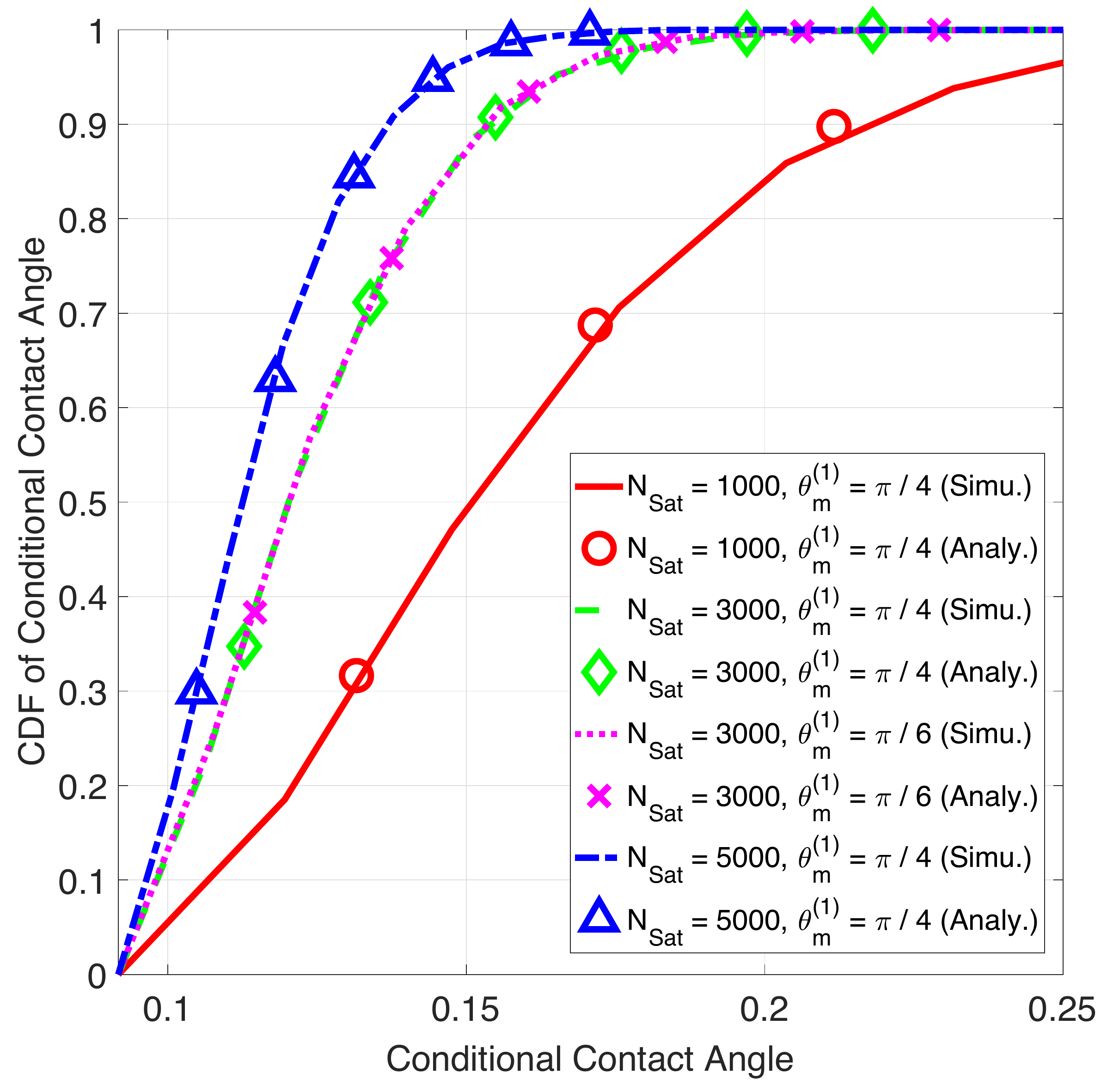}
	\caption{CDF of the conditional contact angle under different number of satellites and the maximum dome angle of the transmitter.}
    \label{fig:Number}
	\vspace{-0.4cm}
\end{figure}

\begin{figure}[t]
	\centering
	\includegraphics[width=0.6\linewidth]{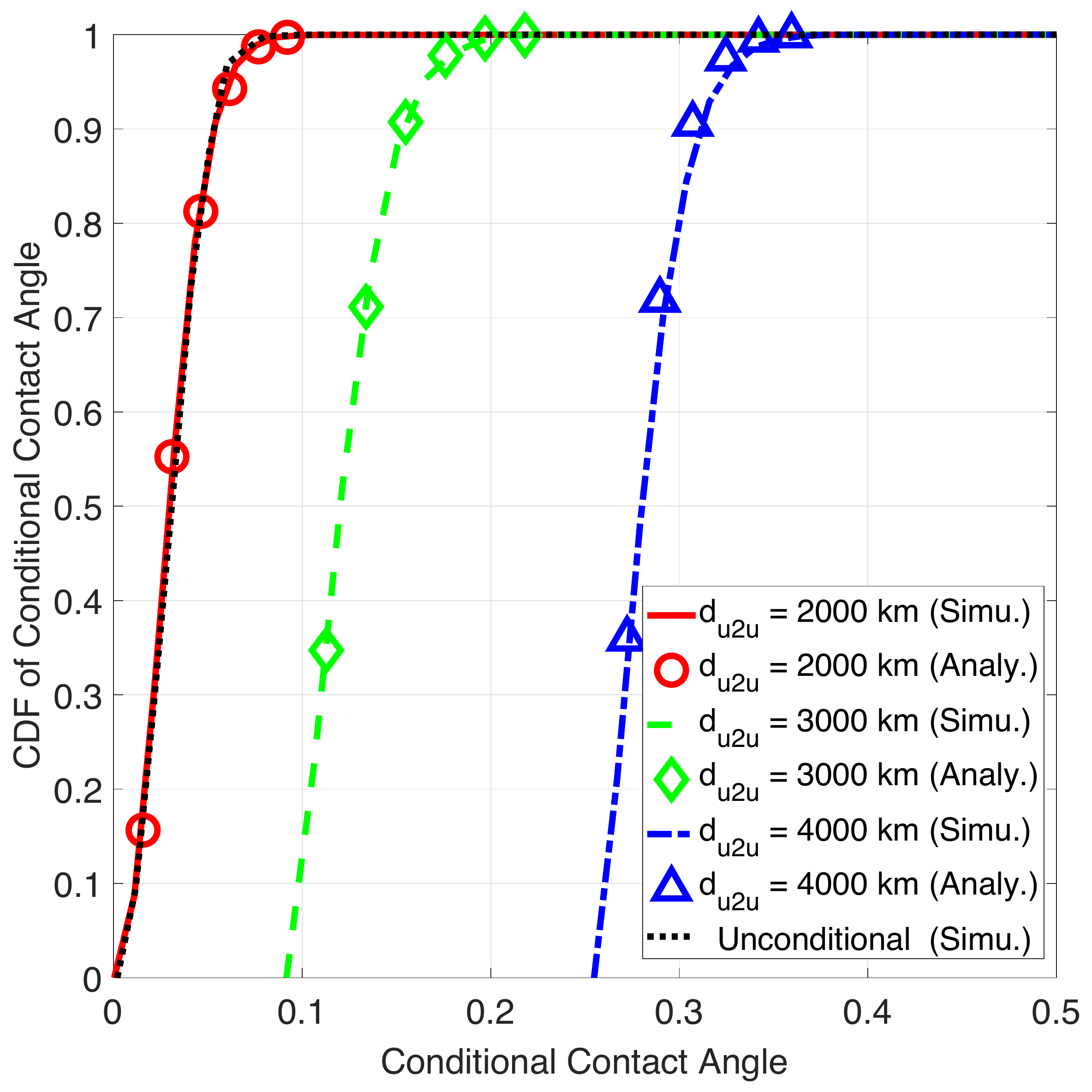}
	\caption{CDF of the conditional contact angle under different distance between the transmitter and receiver.}
    \label{fig:Distance}
	\vspace{-0.4cm}
\end{figure}

In this section, numerical results of the CDF of the conditional contact angle are provided. As shown in Fig.~\ref{fig:Number} and Fig.~\ref{fig:Distance}, the analytical results perfectly match the simulation results, which proves the accuracy of theorem~\ref{CDF theorem}. The height of satellites is fixed at 550km and $R_{\rm{Sat}}=6921\rm{km}$. The maximum dome angle of the receiver is $\theta_m^{(2)}=\frac{\pi}{4}$.

% \begin{figure*}[htbp]
% \begin{minipage}[t]{0.32\linewidth}
% \centering
% \includegraphics[width=0.98\linewidth]{figure3.pdf}
% \caption{CDF of the conditional contact angle under different number of satellites and the maximum dome angle of the transmitter.}
% \label{fig:Number}
% \end{minipage}
% \hfill
% \begin{minipage}[t]{0.32\linewidth}
% \centering
% \includegraphics[width=0.98\linewidth]{figure4.pdf}
% \caption{CDF of the conditional contact angle under different distance between the transmitter and receiver.}
% \label{fig:Distance}
% \end{minipage}
% \hfill
% \begin{minipage}[t]{0.34\linewidth}
% \centering
% \includegraphics[width=0.98\linewidth]{onetwo1.eps}
% \caption{LEO relay outage probability under different distance between the transmitter and receiver and constellation altitudes.}
% \label{fig:Satellite}
% \end{minipage}
% \end{figure*}

% \begin{figure}[h]
% 	\centering
% 	\includegraphics[width=0.65\linewidth]{figure3.pdf}
% 	\caption{CDF of the conditional contact angle under different number of satellites and the maximum dome angle of the transmitter.}
% 	\label{fig:Number}
% 	\vspace{-0.4cm}
% \end{figure}

\par
In Fig.~\ref{fig:Number}, the distributions of the conditional contact angle under the different number of satellites $N_{\rm{Sat}}$ and the maximum dome angle of the transmitter $\theta_m^{(1)}$ are studied. The distance between the ground transmitter and receiver is fixed as $d=3000\rm{km}$.  Reducing $N_{\rm{Sat}}$ causes the CDF curve to move down. Changing $\theta_m^{(1)}$ does not have significant effects on the CDF. 

Fig.~\ref{fig:Distance} describes the influence of the distance between the transmitter and receiver on the distribution of the conditional contact angle. The number of satellites is fixed as $N_{\rm{Sat}}=3000$, and the maximum dome angle of the transmitter is $\theta_m^{(1)}=\frac{\pi}{4}$. As the distance between the transmitter and receiver decreases, the CDF curve is shifted from right to left, with little change in shape. Finally, the conditional contact angle distribution converges to the unconditional contact angle distribution.
\par

% \begin{figure}[h]
% 	\centering
% 	\includegraphics[width=0.65\linewidth]{figure4.pdf}
% 	\caption{CDF of the conditional contact angle under different distance between the transmitter and receiver.}
% 	\label{fig:Distance}
% 	\vspace{-0.4cm}
% \end{figure} 

\begin{figure}[t]
	\centering
	\includegraphics[width=0.6\linewidth]{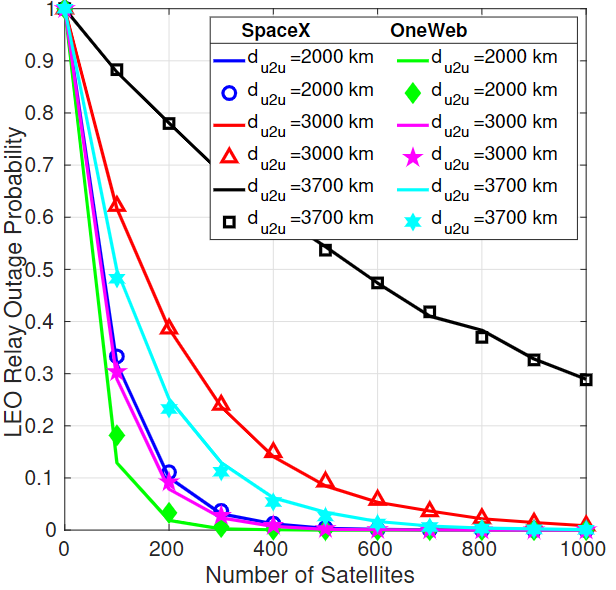}
	\caption{LEO relay outage probability under different distance between the transmitter and receiver and constellation altitudes.}
    \label{fig:Satellite}
	\vspace{-0.4cm}
\end{figure}

In Fig.~\ref{fig:Satellite}, the LEO relay outage probability, $P_e^S \left(d\right) $, is studied under different values for the distance between the transmitter and receiver for companies OneWeb and SpaceX with altitudes $h=1200$km and $h=550$km, respectively. The outage probability gets larger as we increase the distance between the transmitter and receiver. We observe that for the larger altitude of OneWeb, we have less LEO relay outage probability compared with SpaceX for each distance.

% \begin{figure}[t]
% 	\centering
% 	\includegraphics[width=0.65\linewidth]{onetwo1.eps}
% 	\caption{LEO relay outage probability under different distance between the transmitter and receiver and constellation altitudes.}
% 	\label{fig:Satellite}
% 	\vspace{-0.4cm}
% \end{figure}

\section{Conclusion}
In this letter, we derived the approximate conditional contact angle distribution based on the stochastic geometry framework. Three Potential applications of conditional contact angles are further given. Finally, we provide the numerical results about the influence of the number of satellites, the distance between the transmitter and receiver on the conditional contact angle. For the lower altitude of the satellite constellation, it is better to keep the distance less than 3000km to ensure that there are available satellites between the transmitter and receiver for a single-hop transmission.

\appendices
\section{Proof of Lemma~\ref{area}}\label{app:area}
As shown in Fig.~\ref{fig:Decomposition diagrams of the geometry combined by a cone and cap}, the middle part is a geometry composed of a spherical cap and a cone. The right (dark blue) part of the spherical cap in the geometry corresponds to the dome angle $\theta_o$. According to the description in the lemma, the area of the dark blue part is the desired $S\left(\theta_d,\theta_o\right)$. Divide the dome angle $\theta_o$ into infinitesimal $\Delta \theta_o$. The dark blue area can be divided into numerous arcs. $S\left(\theta_d,\theta_o\right)$ is calculated by multiplying the sum of these arcs by $\Delta \theta_o$. Project the spherical cap on the circle in the upper right corner of Fig.~\ref{fig:Decomposition diagrams of the geometry combined by a cone and cap}. The relationship between the arc length $l_{\rm{Arc}}$ and the chord length obtained by projection is given by,
\begin{equation}
    l_{\rm{Arc}}=2{R_{\rm{Sat}}}\arcsin\frac{l_{\perp}}{R_{\rm{Sat}}},
\end{equation}
where $l_{\perp}$ is half of the chord length. Since there is a one-to-one mapping between chord length and arc length, $S\left(\theta_d,\theta_o\right)$ can be obtained by integrating the region on the right side of the circle. As shown in the upper right part of Fig.~\ref{fig:Decomposition diagrams of the geometry combined by a cone and cap}, we choose to integrate in the direction perpendicular to the chord. Easy to know that
\begin{equation}\label{Rcone}
    R_{\rm{Cone}}^2=l_{\varparallel}^2 + l_{\perp}^2=R_{\rm{Sat}}\sin\left(\frac{\theta_d}{2}\right).
\end{equation}
where $R_{\rm{Cone}}$ is the radius of the projected circle. Set the center of the circle as the origin. The upper bound of the integral is $R_{\rm{Cone}}$, and the lower bound is given by,
\begin{equation}\label{lower bound}
    \tan\left(\frac{\theta_d}{2}-\theta_o\right)=\frac{l_{\varparallel}^{\rm{low}}}{R_{\rm{Sat}}\cos\frac{\theta_d}{2}}.
\end{equation}
The integral of $S\left(\theta_d,\theta_o\right)$ is calculated as follows,
\begin{equation}\label{intergral of S}
\begin{split}
    S\left(\theta_d,\theta_o\right)&=\int_{l_{\varparallel}^{\rm{low}}}^{R_{\rm{Cone}}}l_{\rm{Arc}}\left(l\right)\mathrm{d}l\\
    &=\int_{l_{\varparallel}^{\rm{low}}}^{R_{\rm{Cone}}}2{R_{\rm{Sat}}}\arcsin\left(\frac{l_{\perp}}{R_{\rm{Sat}}}\right)\mathrm{d}l\\
    &=\int_{l_{\varparallel}^{\rm{low}}}^{R_{\rm{Cone}}}2{R_{\rm{Sat}}}\arcsin\left(\frac{\sqrt{{R_{\rm{Cone}}}^2-l^2}}{R_{\rm{Sat}}}\right)\mathrm{d}l.
\end{split}
\end{equation}
Substitute (\ref{Rcone}) and (\ref{lower bound}) into (\ref{intergral of S}), the final result of the lemma is obtained. In addition, $l_{\varparallel}^{\rm{low}}$ might be less than 0. It's guaranteed that the lower bound of the integral is always less than the upper bound because,
\begin{equation}
    {\cos\frac{\theta_d}{2}}\tan\left(\frac{\theta_d}{2}-\theta_o\right)<{\cos\frac{\theta_d}{2}}\tan\left(\frac{\theta_d}{2}\right)
    =\sin\left(\frac{\theta_d}{2}\right).
\end{equation}

\section{Proof of Theorem~\ref{CDF of conditional contact angle}}\label{app:CDF of conditional contact angle}
By definition, the CDF of the conditional contact angle can be expressed as,
\begin{equation}\label{CDF - 1}
\begin{split}
    F_{\theta_c}\left(\theta \right) & = 1 - {\mathbb{P}}\left[ \theta_c > \theta \right] = 1 - \mathbb{P}\left[ {\mathcal{N}\left( \mathcal{A}_o \right) = 0} \right] \\
    &\overset{(a)}{=} 1 - \left( 1 - \frac{S\left(\frac{1}{2}\theta_d^{(1)},\theta_o^{(1)}\right)+S\left(\frac{1}{2}\theta_d^{(2)},\theta_o^{(2)}\right)}{4 \pi r^2} \right)^{N_{\rm{Sat}}},
\end{split}
\end{equation}
where $\mathcal{N}\left( \mathcal{A}_o \right)$ counts the number of the satellites in the overlap region $\mathcal{A}_o$. For a homogeneous BPP, the probability of the satellite locates in $\mathcal{A}_o$ is equal to the ratio of the area of $\mathcal{A}_o$ to the total surface area of the sphere. As shown in the bottom of  Fig.~\ref{fig:Satellite-relayed communication model},  $\mathcal{A}_o$ is divided into two parts, with dome angles $\theta_o^{(1)}$ and $\theta_o^{(2)}$. In step (a), the area of $\mathcal{A}_o$ is equal to the sum of $S\left(\frac{1}{2}\theta_d^{(1)},\theta_o^{(1)}\right)$ and $S\left(\frac{1}{2}\theta_d^{(2)},\theta_o^{(2)}\right)$, which are defined in (\ref{S_xitad_xitao}).
\par
When dome angles $\theta_o^{(1)}$, $\theta_o^{(2)}$ and the distance between the transmitter and receiver $d$ are given, $\theta_o^{(1)}$ and $\theta_o^{(2)}$ can be represented by them. From the relationship between these dome angles,
\begin{equation}\label{CDF - 2}
\begin{split}
    \frac{\theta_d^{(1)}}{2}+\frac{\theta_d^{(2)}}{2}-\theta_o^{(1)}-\theta_o^{(2)}=2\arcsin\left(\frac{d}{2R_\oplus}\right).
\end{split}
\end{equation}
Since two right triangles share the red cutting line, the following equation can be obtained,
\begin{equation}\label{CDF - 3}
\begin{split}
    \frac{R_{\rm{Sat}}\cos\left(\frac{1}{2}\theta_d^{(1)}\right)}{\cos\left(\frac{1}{2}\theta_d^{(1)}-\theta_o^{(1)}\right)}=\frac{R_{\rm{Sat}}\cos\left(\frac{1}{2}\theta_d^{(2)}\right)}{\cos\left(\frac{1}{2}\theta_d^{(2)}-\theta_o^{(2)}\right)}
\end{split}
\end{equation}
Combine (\ref{CDF - 2}) and (\ref{CDF - 3}), $\theta_o^{(1)}$ and $\theta_o^{(2)}$ can be derived theoretically. However, it isn't easy to obtain an analytical solution for the two dome angles for this system of trigonometric equations. So we approximate the cosine function by a second-order Taylor expansion,
\begin{equation} \label{CDF - 4}
\cos\left(\frac{\theta_d}{2}-\theta_o\right) \approx 1 - \left(\frac{1}{2}\theta_d - \theta_o \right)^2,
\end{equation}
Substitute (\ref{CDF - 4}) into (\ref{CDF - 3}), we get
\begin{equation} \label{CDF - 5}
    \frac{\cos\left(\frac{1}{2}\theta_d^{(1)}\right)}{1 - \left(\frac{1}{2}\theta_d^{(1)} - \theta_o^{(1)} \right)^2} = \frac{\cos\left(\frac{1}{2}\theta_d^{(2)}\right)}{1 - \left(\frac{1}{2}\theta_d^{(2)} - \theta_o^{(2)} \right)^2}.
\end{equation}
Combine (\ref{CDF - 2}) and (\ref{CDF - 5}), the approximate solution is expressed as,
\begin{equation}
\begin{split}
    \theta_o^{(1)}\left(\theta_d^{(1)}\right) &\approx \frac{1}{2}\theta_d^{(1)} - \frac{ac-\sqrt{2a^2-4ab+2b^2+abc^2}}{a-b}, \\
    \theta_o^{(2)}\left(\theta_d^{(2)}\right) &\approx \frac{1}{2}\theta_d^{(2)} - \frac{-bc+\sqrt{2a^2-4ab+2b^2+abc^2}}{a-b},
\end{split}
\end{equation}

\balance %add on the top of the last page

where $a$, $b$ and $c$ are defined as follows,
\begin{equation}
\begin{split}
    a = \cos\left(\frac{1}{2}\theta_d^{(1)}\right), \ b = \cos\left(\frac{1}{2}\theta_d^{(2)}\right),\  c = 2\arcsin\left(\frac{d}{2R_{\oplus}}\right).
\end{split}
\end{equation}
In this case, $\theta_d^{(2)}$ is fixed as a constant $\theta_m^{(2)}$, while $\theta_d^{(1)}$ is twice the conditional contact angle $\theta_c$. In formula (\ref{CDF - 1}), substitute $\theta_m^{(2)}$ into $\theta_d^{(2)}$ and $2\theta_c$ into $\theta_d^{(1)}$ to get the final result. 
\par
The remaining problem is to determine the range of $\theta_c$. The relay satellite must be within the reliable communication range of the transmitter, so $0 \leq \theta_c \leq \theta_m^{(1)}$ is required. To ensure that the two caps intersect,
\begin{equation}
    2\theta_c+\theta_{m}^{(2)} \geq 4\arcsin \left( \frac{d}{2R_{\oplus}} \right).
\end{equation}
However, when $\theta_c$ is too large, the transmitter's cap may contain the receiver's cap, and increasing $\theta_c$ further is meaningless. Therefore, we have,
\begin{equation}
    2\theta_c \leq 4 \arcsin \left( \frac{d}{2R_{\oplus}} \right)+\theta_{m}^{(2)}.
\end{equation}
In addition, to ensure that the relay satellite is always within the receiver's line of sight, the following inequalities always need to be satisfied,
\begin{equation}
    \theta_m^{(i)}<2\arcsin{\frac{d}{2R_{\oplus}}},\ \ i=1,2.
\end{equation}

\bibliographystyle{IEEEtran}
\bibliography{references}

\end{document}